\documentclass{llncs}

\usepackage{amsmath, amssymb, stmaryrd, wasysym, xspace}
\usepackage{graphicx}

\def\mat#1{\ensuremath{#1}\xspace}

\def\AR#1{\begin{align}#1\end{align}}

\def\EQ#1{\begin{equation}#1\end{equation}}
\def\se#1{Sec.~\ref{#1}}

\def\fig#1{Fig.~\ref{#1}}

\def\ens#1{\{#1\}}

\newcommand{\ket}[1]{\mbox{$|#1\rangle$}}
\newcommand{\gket}{\ket{\psi}}

\newcommand{\ketbra}[2]{\mbox{$|#1\rangle\langle#2|$}}

\newcommand{\proj}[1]{\ketbra{#1}{#1}}

\def\qc{QC_{\cC}}

\def\tM#1#2#3#4{M_{#3#4}^{#1,#2}}

\def\cz#1#2{Z_{#1}^{#2}}
\def\cx#1#2{X_{#1}^{#2}}

\def\et#1#2{E_{#1#2}}

\def\nr{ \;|\; }
\def\given{\, \| \,}

\def\qc{\mathtt{qc}} 
\def\cc{\mathtt{c}} 
\def\qci{\mathtt{qc?}} 
\def\cci{\mathtt{c?}} 
\def\qco{\mathtt{qc!}} 
\def\cco{\mathtt{c!}} 
\def\init#1{\mrm{init(}#1\mrm{)}}
\def\fin#1{\mrm{fin(}#1\mrm{)}}

\def\Lrar{\Longrightarrow}

\def\Slar#1{\stackrel{#1}{\Lrar}}

\newcommand{\Ent}{\vDash}

\def\emptyset{\varnothing}

\def\alw{\Box}
\def\eve{\Diamond}
\def\nex{\fullmoon}

\def\mrm{\mathrm}
 
\def\mbf{\mathbf}

\def\mbb{\mathbb}

\def\ga{\gamma}
\def\Ga{\Gamma}

\def\sig{\mat{\sigma}}
\def\ta{\theta}

\newcommand{\bi}{\ensuremath{\mathbf{i}}\xspace}

\newcommand{\bo}{\ensuremath{\mathbf{o}}\xspace}

\newcommand{\bA}{\ensuremath{\mathbf{A}}\xspace}
\newcommand{\bB}{\ensuremath{\mathbf{B}}\xspace}

\newcommand{\cC}{\ensuremath{\mathcal{C}}}

\newcommand{\cE}{\ensuremath{\mathcal{E}}}

\newcommand{\cN}{\ensuremath{\mathcal{N}}}

\begin{document}

\title{Reasoning about Quantum Knowledge}

\author{Ellie D'Hondt\inst{1} \and Prakash Panangaden\inst{2}}

\institute{Vrije Universiteit Brussel, Belgium \\ \email{Ellie.DHondt@vub.ac.be}\\ \and McGill University,  Canada \\\email{prakash@cs.mcgill.ca}}
\maketitle

\begin{abstract}
We construct a formal framework  for investigating epistemic and temporal notions in the context of distributed quantum computation. While we rely on structures developed in \cite{Danos05b}, we stress that our notion of quantum knowledge makes sense more generally in any agent-based model for distributed quantum systems. Several arguments are given to support our view that an agent's possibility relation should not be based on the reduced density matrix, but rather on local classical states and local quantum operations. In this way, we are able to analyse distributed primitives such as superdense coding and teleportation, obtaining interesting conclusions as to how the knowledge of individual agents evolves. We show explicitly that the knowledge transfer in teleportation is essentially classical, in that eventually, the receiving agent knows that its state is equal to the initial state of the sender. The relevant epistemic statements for teleportation deal with this correlation rather than with the actual quantum state, which is unknown throughout the protocol.
\end{abstract}

\section{Introduction}\label{cdk}

The idea of developing formal models to reason about knowledge has proved to be very useful for distributed systems \cite{Hintikka62,Halpern95,Fagin95}. Epistemic logic provides a natural framework for expressing the knowledge of agents in a network, allowing one to make quite complex statements about what agents know, what they know that other agents know, and so on. Moreover, combining epistemic with temporal logic, one can investigate how knowledge evolves over time in distributed protocols, which is useful both for program analysis as well as formal verification. 

The standard approach to knowledge representation in multi-agent systems is based on the \emph{possible worlds} model. The idea is that there exists a set of worlds such that an agent may consider several of these to be possible. An agent \emph{knows} a fact if it is true in all the worlds it considers possible; this is expressed by epistemic modal operators acting on some basic set of propositions. The flexibility of this approach lies in the fact that there are many ways in which one can specify possibility relations. In a distributed system, worlds correspond to global configurations occurring in a particular protocol, and \emph{possible} worlds are determined by an equivalence relation over these configurations. Typically, global network configurations are considered equivalent by an agent if its local state in these configurations is identical. 

Quantum computation is a field of research that is rapidly acquiring a place as a significant topic in computer science~\cite{Nielsen00}. Logic-based investigations in quantum computation are relatively recent and few. Recently there have been some endeavours in describing quantum programs in terms of predicate transformers ~\cite{Baltag04,Baltag05,Vandermeyden03}. These frameworks, however, aim at modelling traditional algorithms that establish an input-output relation, a point of view which is not appropriate for distributed computations. A first attempt to define knowledge for quantum distributed systems is found in \cite{Vandermeyden03b}. Therein, two different notions of knowledge are defined. First, an agent $i$ can \emph{classically} know a formula $\ta$ to hold, denoted $K_{i}^{c}\ta$; in this case the possibility relation is based on equality of local classical states. Second,  an agent can \emph{quantumly} know a formula to hold, denoted $K_{i}^{q}\ta$. For the latter, the possibility relation is based on equality of reduced density matrices for that agent. The authors argue that $K_{i}^{q}$ is an information-theoretic idealisation of knowledge, in that the reduced density matrix embodies what an agent, in principle, could determine from its local quantum state. However, there are two main problems with this approach. The first is that one cannot assume that the reduced density matrix is always known, because in quantum mechanics, observing a state alters it irreversibly. So, quantum knowledge does not consist of possession of a quantum state:  it is not because an agent has a qubit in its lab that the agent knows anything about it. Indeed, consider the situation where a qubit has just been sent from \bA to \bB. Then \bB knows nothing about its newly acquired qubit -- it is possible, even,  that \bA knows more about it than \bB does. The second problem with the above approach is that one loses information on correlations between agents by considering only the reduced density matrix, a crucial ingredient in distributed quantum primitives.

What we need is a proper notion of quantum knowledge, which captures the information an agent can obtain about its quantum state. This includes the following ingredients: first, an agent knows states that it has prepared; second, an agent knows a state when it has just measured it; and third, an agent may obtain knowledge by classical communication of one of the above. While knowledge of preparation states is automatically contained in the description of the protocol, our notion of equivalence precisely captures the latter two items. As we shall see below, in doing this we find a similar notion as $K_{i}^{c}\ta$. Our main argument, then, is that there is no such thing as quantum knowledge in the sense of $K_{i}^{q}\ta$; rather quantum knowledge is about classically knowing facts about quantum systems.

The structure of this paper is as follows. In \se{qdk} we construct a framework for reasoning about knowledge in quantum distributed systems. Next, we investigate the important distributed primitives of superdense coding and teleportation in our epistemic framework in \se{apps}, investigating how agents' knowledge is updated as each protocol proceeds. We conclude in \se{summary5}. 

This paper assumes some familiarity with quantum computation -- for the reader not familiar with the domain, we refer to the excellent ~\cite{Nielsen00}. The present paper is also a continuation of earlier work by the authors ~\cite{Danos04b,Danos05b}. However, most of the material presented here can be understood independently of the latter.

\section{Knowledge in Quantum Networks}\label{qdk}

In this section, we develop the notion of knowledge for distributed quantum systems. The equivalence relation for agents, on the basis of which quantum knowledge is defined, is established in \se{knowledge}. Next, temporal operators are defined in \se{time}, where we also briefly discuss how temporal and epistemic operators combine.

We phrase our results below in the context of  \emph{quantum networks}, an agent-based model for distributed quantum computation elaborated in \cite{Danos05b}. We stress, however, that our notion of quantum knowledge is model-independent. That is to say, any agent-based model for distributed quantum computation would benefit from quantum knowledge as defined below, or slight adaptations thereof. Due to space limitations, only a short overview is given here; for more detailed explanations, we refer the reader to \cite{Danos05b,DHondt05}. 

A \emph{network of agents} $\cN$ is defined by a set of concurrently acting agents together with a shared quantum state, that is
\EQ{
\cN= \bA_{1}:Q_{1}.\cE_{1}\nr\dots\nr\bA_{m}:Q_{m}.\cE_{m}\given \sig=|_{i}\bA_{i}(\bi_{i},\bo_{i}):Q_{i}.\cE_{i}\given \sig \text{  ,}
}
where $\sig$ is the network quantum state, $|$ denotes parallel composition, and for all $i$, $\bA_{i}$ is an agent with local qubits $Q_{i}$ and event sequence $\cE_{i}$. The network state \sig in the definition is the initial entanglement resource which is distributed among agents. Local quantum inputs are added to the network state \sig during initialisation; in this way we keep initial shared entanglement as a first-class primitive in our model. Note that agents in a network need to have different names, since they correspond to different parties that make up the distributed system. In other words, concurrency comes \emph{only} from distribution; we do not consider parallel composition of processes in the context of one party.  Events consist of local quantum operations $A$, classical communication $\cci$ and $\cco$, and quantum communication $\qci$ and $\qco$. Quantum operations are denoted in the style of \cite{Danos04b}, that is we have entanglement operators $E$, measurements $M$ and Pauli corrections $X$ and $Z$. All of this is much clarified in the applications in \se{apps}.

A network determines a set of configurations  $\cC_{\cN}$ that can potentially occur during execution of $\cN$. Configurations are written
\EQ{ 
C=\ket{\sig}, |_{i}\Ga_{i}, \bA_{i}:Q_{i}.\cE_{i}\text{  ,}
}
where $\Ga_{i}$ is each agent's local (classical) state, which is where measurement outcomes and classical messages are stored. $\cC_{\cN}$ consists of all configurations encountered in those paths a protocol can take. More formally, $\cC_{\cN}$ is obtained by following the rules for the small-step operational semantics of networks, denoted by transitions $\Lrar$ and elaborated in \cite{Danos05b}.

Before we can actually define modal operators for knowledge or time, we need to clarify what the propositions are that these act upon. It is not our intention to define a full-fledged language for primitive propositions; rather, we define these abstractly.  An \emph{interpretation} of $\cN$ is a truth-value assignment for configurations in  $\cC_{\cN}$ for some basic set of  primitive propositions $\ta$. Writing $I(C,\ta)$ for the interpretation of fact $\ta$ in configuration $C$, we then have,
\EQ{
C,\cN \Ent \ta \iff I(C,\ta)=\mrm{true}\text{  .}
}
The primitive propositions considered usually depend on the network under study, and are specified individually for each application encountered below. Composite formulas can be constructed from primitive propositions and the logical connectives $\land$, $\lor$ and $\neg$ in the usual way. However, the formulas encountered in the applications below are usually about equality. For example, $\ta$ may be of the form $x=v$, meaning that the classical variable $x$ has the value $v$, or $q_{1}=q_{2}$, meaning that the states of qubits $q_{1}$ and $q_{2}$ are identical. We also allow functions \emph{init} and \emph{fin} for taking the initial and final values of a variable or quantum state. These formulas are currently defined in an ad-hoc manner.

\subsection{Knowledge}\label{knowledge}

In order to define quantum knowledge, we need to define an equivalence relation on configurations for each of the agents, embodying what an agent knows about the global configuration from its own information only. We deliberately do not say \emph{local} information here, as, via the network preparation, an agent may also have non-local information, under the form of correlations, at its disposal. By considering only configurations in $\cC_{\cN}$ we model that agents know which protocol they are executing.

In a quantum network, each agent's equivalence relation has to reflect what an agent knows about the network state, the execution of the protocol and the results of measurements. All classical information an agent has is stored in its local state $\Ga$; this includes classical input values, measurement outcomes, and classical values passed on by other agents. Just like in classical distributed systems, an agent can certainly differentiate configurations for which the local state is different. As for quantum information, an agent knows which qubits it owns, what local operations it applies on these qubits, and, moreover, what (non-local) preparation state it starts out with, i.e. what entanglement it shares with other agents initially. It can also have information on its local quantum inputs, though this is not necessarily so, as we have explained in the above. All of the above information is in fact captured by an agent's event sequence in a particular configuration, together with its local state. Therefore, we obtain the following definition.

\begin{definition}
Given a network $\cN$ and configurations $C=\sig;\, |_{i}\Ga_{i}, \bA_{i}:Q_{i}.\cE_{i}$ and $C'=\sig';\, |_{i}\Ga'_{i}, \bA_{i}:Q'_{i}.\cE'_{i}$ in $\cC_{\cN}$, we say that agent $\bA_{i}$ considers $C$ and $C'$ to be \emph{equivalent}, denoted $C \sim_{i}C'$, if $\Ga_{i}=\Ga'_{i}$ and $\cE_{i}=\cE'_{i}$.
For each agent $\bA_{i}$ the relation $\sim_{i}$ is an equivalence relation on $\cC_{\cN}$, called the \emph{possibility relation} of $\bA_{i}$.
\label{equivalence}
\end{definition}

Via possibility relations we can now define what it means for an agent $\bA_{i}$ to know a fact $\ta$ in a configuration $C$ in the usual way,
\EQ{
C,\cN \Ent K_{i}\ta \iff \forall C'\sim_{i}C : C'\Ent \ta \text{  .}
\label{k}
}

Our choice of equivalence embodies that agents cannot distinguish configurations if they only differ in that other agents have applied local operations to their qubits; neither can they if other agents have exchanged messages with each other. While the global network state does change as a result of local operations, an agent not executing these has no knowledge of this, and no way of obtaining it. This is precisely what we capture with the relation $\sim_{i}$. 

Special attention needs to be given to the matter of quantum inputs. Agents distinguish configurations corresponding to different values of their classical input via their local state, in which these input values are stored. Essentially, for each set of possible input values there is a group of corresponding configurations in $\cC_{\cN}$. However, this is not something we can do for quantum inputs, since these occupy a continuous space. Hence we choose to let configurations be parameterised by these inputs, writing $C(\gket)$ whenever we want to stress this. But then what about an agent's possibility relation? Basically, either a quantum input is known, in which case it is just a local preparation state such that there is only one possible initial configuration. If a quantum input for agent $\bA$ is truly arbitrary, or the agent knows nothing about it -- as is the case for teleportation -- then all values of $\gket$, and hence all configurations in the set $\ens{C(\gket),\gket \in I_{\bA}}$, are considered equivalent by \bA. If \bA does know some properties of its input, then we model this by only allowing a certain set of input states. We do not explicitly mention the equivalence related to unknown quantum inputs in the examples below, for the simple reason that we are interested only in logical statements that hold for \emph{all} quantum inputs. That is, we compare only configurations resulting from the same quantum inputs, and derive knowledge-related statements that are independent of this input. Nevertheless, whenever a configuration $C(\gket)$ is written, it should be interpreted as a \emph{set} of states, all considered equivalent by all agents of the network.

From this one can construct more complicated statements, such as for example $C,\cN \Ent K_{\bA}K_{\bB}\ta$ for ``agent \bA knows that agent \bB knows that $\ta$ holds in configuration $C$'.

\subsection{Time}\label{time}

One typically also wants to investigate how knowledge \emph{evolves} during a computation, for example due to communication between agents. Thus, one also needs a proper formalisation of  \emph{time}. This is usually done by allowing a set of \emph{temporal} modal operations, operating on the same set of propositions. The area of temporal logics is itself an active field of research, with applications in virtually all aspects of concurrent program design; for an overview see for example \cite{Emerson90}.

We use the approach of  \emph{computational tree logic} (CTL) to formalise time-related logical statements, providing state as well as path modal operators. The reason for this is that, due to the fact that quantum networks typically have a branching structure, we need to be able to express statements concerning \emph{all} paths as well as those pertaining to \emph{some} paths. Typically, we want to say things such as ``for all paths, agent \bA always knows $\ta$ '', or ``there exists a path for which \bA eventually knows $\ta$''. We can of course express this by placing restrictions on the paths we are considering in a particular statement -- this is, in fact, precisely what we do in the definition of modal path operators. Introducing these is more appealing since in this way we can abstract away from actual path definitions, which are determined by the formal semantics for networks elaborated in ~\cite{Danos05b}, and denoted abstractly as $\Lrar$ below. 

Concretely, we introduce the traditional temporal state operators $\alw$ (``always'') and $\eve$ (``eventually'') into our model, and combine these with the path operators $A$ (``for all paths'') and $E$ (``there exists a path''), as follows\footnote{$\Slar{\ga}$ is the closure of the small-step transition relation $\Lrar$ mentioned above. That is, we have $C\Slar{\ga}C'$ if $C'$ can be reached form $C$ by a series of consecutive small-step transitions, specified by the path $\ga$.}

\AR{
C,\cN \Ent A\alw \ta &\iff \forall \ga, \forall C' \text{  with  }    C\Slar{\ga} C': C' \Ent \ta \label{aalw}\\
C,\cN \Ent E \alw \ta &\iff \exists \ga, \forall C'\text{  with  }  C\Slar{\ga}C': C' \Ent \ta \\
C,\cN \Ent A \eve \ta &\iff \forall \ga, \exists C'\text{  with  }  C\Slar{\ga} C': C' \Ent \ta\\
C,\cN \Ent E \eve \ta &\iff  \exists \ga, \exists C'\text{  with  }  C\Slar{\ga} C' : C' \Ent \ta \text{  .}
\label{enex}
}

Obviously, we have that any formula with $A$ implies the corresponding one with $E$, and likewise any formula with $\alw$ implies the corresponding ones with $\eve$ and $\nex$.

When investigating knowledge issues in a distributed system, one naturally arrives at situations where one needs to describe formally how knowledge evolves as the computation proceeds. This can be done adequately by combining knowledge operators $K_{i}$ with the temporal operators defined above. As usual, one needs to proceed with caution when doing this, since it is not always intuitively clear what the meaning of each of these different combinations is. For example, it is generally \emph{not} the case that the formula $A\alw K_{i}\ta$ is equivalent to $K_{i}A\alw \ta$. Typically, we want to prove things that are eventually known by an agent, no matter what branch the protocol follows; this is embodied by the former. 

\section{Applications}\label{apps}

With epistemic and temporal notions for quantum networks in place, we are ready to evaluate the distributed primitives superdense coding~\cite{Bennett92} and teleportation~\cite{Bennett93} from a knowledge-based perspective. That is, instead of investigating how the global network evolves by deriving a network's semantics, we now use this semantics, or rather, the configurations encountered therein, to analyse how the knowledge of individual agents evolves. We start with superdense coding, which is simpler to analyse because it is deterministic and does not depend on quantum inputs. We move on to teleportation in \se{tpk}. We note that an analysis of the quantum leader election protocol~\cite{DHondt05b} was also carried out in \cite{DHondt05}.

\subsection{Superdense Coding}

The aim of superdense coding is to transmit two classical bits from one party to the other with the aid of one entangled qubit pair or ebit. The network for this task is defined as follows,
\EQ{
SC=\bA:\ens{1}.[(\qco 1)\cx {1}{x_{2}}\cz {1}{x_{1}}]\nr\bB:\ens{2}.[ \tM 0012 (\qci 1)]\given \et 12 \text{  ,} 
\label{dcn}
}
Here $x_{1}x_{2}$ are \bA's classical inputs, subscripts stand for qubits on which events operate, $X$ and $Z$ are Pauli operations, $\qco$ and $\qci$ stand for a quantum rendezvous, $ \tM 0012$ is a Bell measurement on qubits 1 and 2, and $\et 12$ is an ebit. In the first step of the protocol Alice transforms her half of the entangled pair, in a different way for each of the four possible classical inputs. Next, she sends Bob her qubit, who then measures the entangled pair. At the end of the protocol the measurement outcomes, denoted $s_{1}$ and $s_{2}$, are equal to \bA's inputs. 

The configurations in $\cC_{SC}$ are the following~\cite{DHondt05},
\AR{
C_{1}^{j_{1}j_{2}}&=  \et 12;[x_{1},x_{2}\mapsto j_{1},j_{2}],\bA:\ens{1}.[(\qc !1)\cx {1}{x_{2}}\cz {1}{x_{1}}]\nr\emptyset,\bB:\ens{2}. [ \tM 0012 (\qci 1)]\notag\\
C_{2}^{j_{1}j_{2}}&=\cx {1}{x_{2}}\cz {1}{x_{1}}\et 12;[x_{1},x_{2}\mapsto j_{1},j_{2}],\bA:\ens{1}.(\qco1)\nr\emptyset,\bB:\ens{2}. [ \tM 0012 (\qci 1)]\notag\\
C_{3}^{j_{1}j_{2}}&=\cx {1}{x_{2}}\cz {1}{x_{1}}\et 12;[x_{1},x_{2}\mapsto j_{1},j_{2}],\bA \nr\emptyset,\bB:\ens{1,2}. \tM 0012\notag\\
C_{4}^{j_{1}j_{2}}&=\mbf{0};[x_{1},x_{2}\mapsto j_{1},j_{2}]\bA\nr [s_{1},s_{2}\mapsto j_{1},j_{2}],\bB\text{  ,}\notag
}
where $j_{1}j_{2}$ is equal to the input values 00,01,10 or 11. 

\begin{figure}
\begin{center}
\scalebox{.50}{\includegraphics{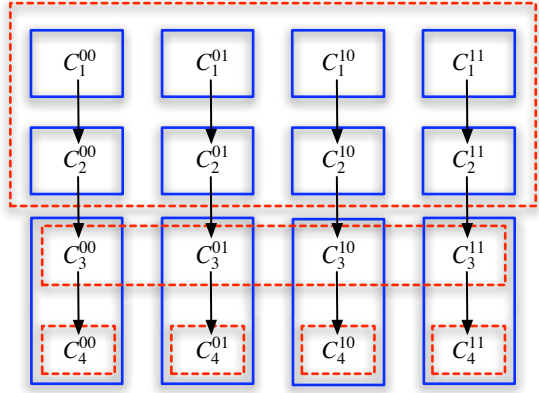}}
\caption[ ]{Possibility relations for the superdense coding network.}
\label{sdequivalence}
\end{center}
\end{figure}

The equivalence relation for both of the agents for configurations in $\cC_{SC}$ is represented in \fig{sdequivalence}, with arrows for computation paths, boxes for \bA's equivalence classes and dashed boxes for \bB's equivalence classes. Obviously, $\bA$ distinguishes the 4 possible configurations at each time step -- we refer to this below as \emph{horizontally} -- because \bA's local state $[x_{1},x_{2}\mapsto j_{1},j_{2}]$ is different for each input value.  \emph{Vertically}, that is with respect to the evolution of time, configurations at the first three steps differ because \bA's event sequence has changed. However, we find that configurations at the third and fourth level are equivalent for \bA, since from between both steps \bB has applied a local operation, which is not observable by \bA. 

The possibility relation for \bB is quite different. We find that that all configurations occurring at the first two steps are considered equivalent by \bB. Furthermore, all configurations $C_{3}$ are equivalent to each other, though they are not equivalent to the previous ones because the event sequence of \bB has changed. Configurations $C_{4}$ differ from the previous ones because here \bB applies a local operation, and furthermore, here \bB finally distinguishes states horizontally via its local state $[s_{1},s_{2}\mapsto j_{1},j_{2}]$. 

The possibility relations of both agents allow us to derive several epistemic statements. First of all, however, let us note that the SC network is correct, since we have
\EQ{
\forall j_{1},j_{2}: C_{1}^{j_{1}j_{2}}, SC \Ent A\eve(s_{1}s_{2}=j_{1}j_{2})\text{  ,}
}
or, if we want to stress that this occurs in the last step, we use $C_{3}(j_{1}j_{2}), SC \Ent A\nex (s_{1}s_{2}=j_{1}j_{2})$. Note that, since there is no branching in the protocol, we may replace $A$ by $E$ in the above. 

Next, we trivially have that $C , SC\Ent K_{\bA}(x_{1}x_{2}=j_{1}j_{2})$ for all $C\in\cC_{SC}$, that is, \bA always knows its input values -- in fact, agents always know their own input values in any protocol. We can also state this by saying that for all input values $C_{1}^{j_{1}j_{2}}, SC \Ent A\alw K_{\bA}(x_{1}x_{2}=j_{1}j_{2})$. On the other hand, it is only in the last step that \bB knows \bA's input values, that is 
\EQ{
\forall j_{1},j_{2}: C_{4}^{j_{1}j_{2}}, SC \Ent K_{B}(s_{1}s_{2}= j_{1}j_{2})\text{  ,}
}
while
\EQ{
\forall j_{1},j_{2}, s<4: C_{s}^{j_{1}j_{2}}, SC \Ent \neg K_{B}(s_{1}s_{2}= j_{1}j_{2})\text{  .}
}
Interestingly, \bA never knows that \bB knows \bA's input values eventually, 
\EQ{
\forall j_{1},j_{2}: C_{1}^{j_{1}j_{2}}, SC \Ent \neg A\eve K_{\bA} K_{\bB}(s_{1}s_{2}=j_{1}j_{2})\text{  .}
}
The reason for this is that \bA cannot distinguish between configurations at the last two time steps, that is, \bA does not know whether \bB has applied its local measurement yet, and therefore \bA never knows if \bB knows that $s_{1}s_{2}=j_{1}j_{2}$.  

Other statements that can be made about the SC network, for example one can play around with temporal operators to highlight when exactly the quantum message is sent. However, the essential features of the protocol are captured above.

\subsection{Teleportation}\label{tpk}

The goal of the teleportation network is to transmit a qubit from one party to another with the aid of an ebit and classical resources. The network achieving this is defined as follows,
\EQ{
TP=\bA:\ens{1,2}.[(\cc!s_{2}s_{1}).\tM 0012] \nr \bB:\ens{3}.[\cx 3{x_{2}}\cz 3{x_{1}}.(\cc?x_{2}x_{1})] \given \et 23\text{  ,}
\label{tpn}
}
where  $\cco$ and $\cci$ stand for a classical message rendezvous.
In the first step of the protocol Alice executes a Bell measurement on her qubits. Next, Alice sends Bob her measurement outcomes, after which Bob applies Pauli corrections to his qubit dependent on these outcomes. The result is that Bob's qubit ends up in the same state as Alice's input qubit. 

In this case, we have branching due to the Bell measurement. Moreover, configurations are parameterised by the quantum input $\gket$. As explained above, we do not explicitly show that configurations for different quantum inputs are equivalent for all agents. This feature is usually expressed by saying that $\gket$ is an unknown quantum state, that is, \bA (nor \bB) know anything about it. We repeat the configurations occurring throughout the execution of the protocol explicitly here, labelling configurations by measurement outcomes obtained in the first step of the computation. 
\AR{
C_{1}(\gket)&=  \gket\et 23;\emptyset,\bA:\ens{1,2}.[(\cc!s_{2}s_{1}).\tM 0012]\nr\emptyset,\bB:\ens{3}.[\cx 3{x_{2}}\cz 3{x_{1}}.(\cc?x_{2}x_{1})]\notag\\
C_{2}^{j_{1}j_{2}}(\gket)&=X^{j_{2}}Z^{j_{1}}\gket; [s_{1},s_{2}\mapsto j_{1},j_{2}],\bA.(\cc!s_{2}s_{1})|\emptyset,\bB:\ens{3}.[\cx 3{x_{2}}\cz 3{x_{1}}.(\cc?x_{2}x_{1})]\notag\\
C_{3}^{j_{1}j_{2}}(\gket)&=X^{j_{2}}Z^{j_{1}}\gket; [s_{1},s_{2}\mapsto j_{1},j_{2}],\bA\nr[x_{1},x_{2}\mapsto j_{1},j_{2}],\bB:\ens{3}.\cx 3{x_{2}}\cz 3{x_{1}})\notag\\
C_{4}^{j_{1}j_{2}}(\gket)&=\gket;[s_{1},s_{2}\mapsto j_{1},j_{2}],\bA\nr [x_{1},x_{2}\mapsto j_{1},j_{2}],\bB:\ens{3}\text{  .}\notag
}

\begin{figure}
\begin{center}
\scalebox{.50}{\includegraphics{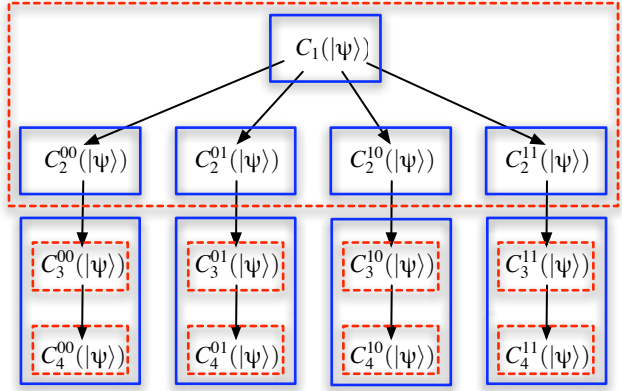}}
\caption{Possibility relations for the teleportation network.}
\label{tpequivalence}
\end{center}
\end{figure}

The equivalence relation for both agents for the set of configurations $\cC_{TP}$ is represented in \fig{tpequivalence}. We find that $C_{1}(\gket)$ is equivalent only to itself for agent \bA --  once more, in effect we have a set $\ens{C_{1}(\gket),\gket \in \mbb C^2}$ of equivalent configurations with respect to $\sim_{\bA}$. After the measurement \bA distinguishes (sets of) configurations horizontally at all time steps via its outcome map. Just as for SC, and for the same reason, \bA considers configurations at the last two steps to equivalent.

Again, the situation for agent \bB is quite different. We find that that configurations at the first two levels are considered to be equivalent, while all other configurations are distinguished, horizontally via \bB's local state, and vertically by the change in \bB's event sequence.

The correctness of the TP network is stated in logical terms as follows,
\EQ{
C_{1},TP\Ent A\eve(\fin{q_{3}}=\init{q_{1}})\text{  ,}
}
where we have left out the parameterisation because the statement holds for all $\gket$.  In other words, the final state of \bB's qubit $q_{3}$ is identical to the initial value of \bA's qubit $q_{1}$\footnote{We refer to the qubit named $q_{i}$ as qubit $i$ in semantical derivations.}. Interestingly, neither of the agents know the actual quantum state at \emph{any} point of the computation, that is
\AR{
C_{1}(\gket),TP &\Ent \neg K_{\bA}(q_{1}=\gket)\land\neg K_{\bB}(q_{1}=\gket)\\
C_{1}(\gket),TP  &\Ent \neg E\eve K_{\bA}(q_{3}=\gket)\land  \neg E\eve K_{\bB}(q_{3}=\gket)\text{  ,}
}
that is to say, initially nobody knows that $q_{1}$ is in the state $\gket$, and there is no future point in the protocol at which either \bA or \bB knows that $q_{3}$ is in the state $\gket$. The basic reason for this is of course that for all input states $\gket$ configurations $C(\gket)$ are considered equivalent by all agents, and therefore they can conclude nothing about properties $\gket$ may have. Apart from statements about classical message passing,  in TP the only knowledge transfer deals with the \emph{correlation} between initial and final states of the network, \emph{not} with the actual form of the quantum input. To be more precise, we have that
\EQ{
C_{1},TP \Ent A\eve K_{\bB}(q_{3}=\init{q_{1}})\text{ ,}
}
since at the last step of the computation \bB knows that it must have the original input state. However, since \bA cannot distinguish the last two time steps, we also have that
\EQ{
C_{1},TP \Ent \neg E\eve K_{\bA}(q_{3}=\init{q_{1}})\text{ .}
}
The latter two statements may seem odd in that we are talking about states that the agents know nothing about. However, even without knowing a state, one may still have information about how it compares with other states. There is nothing strange about this, as this sort of thing happens with classical correlations too. What it does show, however, is that there is no actual quantum knowledge transfer in the TP network -- there was no quantum knowledge about the input to begin with! We can only say something about the relation of the initial to final quantum states.

Note that our analysis is in stark contrast with the one found in \cite{Vandermeyden03b}, which is jointly in terms of $K_{i}^{c}$ and $K_{i}^{q}$. As mentioned above, the latter is based upon equality of reduced density matrices. Next to our objections to this approach mentioned earlier, such an analysis becomes increasingly awkward when applied to the teleportation protocol, since the basis of TP is that the initial state is \emph{unknown}. In fact, the authors themselves note that their analysis leads to difficulties. Concretely, in their framework the conclusion is that initially \bA has quantum knowledge of $\gket$ --  i.e. \bA knows its initial reduced density matrix, which is just $\proj{\psi}$ --  while \bB does not, and that eventually \bB knows the initial state $\gket$, i.e. the same reduced density matrix. However, if Alice teleports a single qubit to Bob she absolutely has not transmitted a continuum of information. Indeed, Bob needs many such qubits to determine, via statistical analysis, which quantum state has been teleported. Moreover, as pointed out by the authors themselves, their notion of knowledge allows \bB to distinguish the four possible network states even \emph{before} \bA has sent the measurement results through, i.e. at the second step of the computation. This is not the case: in fact the classical message passing is \emph{crucial} for the success of the protocol, as without this information Bob's state is given by the maximally mixed state. All these arguments just strengthen our point: analysing teleportation from an epistemic point of view has nothing to do with quantum states, but rather with the relationship between them. Our point is that, although quantum mechanics can be used to transmit information in unexpected ways, there is no such thing as quantum knowledge; it is all classical knowledge, albeit about quantum systems.

\section{Conclusion} \label{summary5}

We have developed a formal framework for investigating epistemic and temporal notions in the context of distributed quantum systems. While we rely on structures developed in prior work, our notion of quantum knowledge makes sense more generally in any agent-based model of quantum networks. Several arguments are given to support our view that an agent's possibility relation should not be based on the reduced density matrix, but rather on local classical states and local quantum operations.  In this way, we are able to analyse distributed primitives from a knowledge-based perspective. Concretely, we investigated superdense coding and teleportation, obtaining interesting conclusions as to how the knowledge of individual agents evolves. We have explicitly shown that the knowledge transfer in teleportation is essentially classical, in that eventually, the receiving agent only knows that its state is equal to the initial state of the sender. The relevant epistemic statements for teleportation deal with this correlation rather than with the actual quantum state, which is unknown throughout the protocol.

\end{document}